\documentclass[%
preprint,
amsmath,amssymb,
aps,
floatfix,
]{revtex4-2}
\usepackage{xcolor}
\usepackage{graphicx}
\usepackage{dcolumn}
\usepackage{bm}
\usepackage{hyperref}

\begin{document}

\title{Cooperative Motions and Topology-Driven Dynamical Arrest in Prime Knots} 

\author{Hyo Jung Park}
\affiliation{Smith College}

\author{Anna Lappala}
\email{lappala@molbio.mgh.harvard.edu}
\affiliation{Massachusetts General Hospital \& Department of Molecular Biology, Harvard University}

\date{\today}
\maketitle

Knots are entangled structures that cannot be untangled without a cut. Topological stability of knots is one of the many examples of their important properties that can be used in information storage and transfer. Knot dynamics is important for understanding general principles of entanglement as knots provide an isolated system where tangles are highly controlled and easily manipulated. To unravel the dynamics of these entangled topological objects, the first step is to identify the dominant motions that are uniquely guided by knot structure and its complexity. We identify and classify motions into three main groups— orthogonal, aligned, and mixed motions, which often act in unison, orchestrating the complex dynamics of knots. The balance between these motions is what creates an identifiable signature for every knot. As knot complexity increases, the carefully orchestrated dynamics is gradually silenced, eventually reaching a state of topologically driven dynamical arrest. Depending on their complexity, knots undergo a transition from nearly stochastic motions to either non-random or even quasiperiodic dynamics before culminating in dynamical arrest. Here, we show for the first time that connectivity alone can lead to a topology-driven dynamical arrest in knots of high complexity. Unexpectedly, we noticed that some knots undergo cooperative motions as they reach higher complexity, uniquely modulating conformational patterns of a given knot. Together, these findings demonstrate a link between topology and dynamics, presenting applications to nanoscale materials.   

\clearpage

\section{Introduction}\label{sec: intro}

\begin{figure}[ht]
    \centering
    \includegraphics[width=0.7\linewidth]{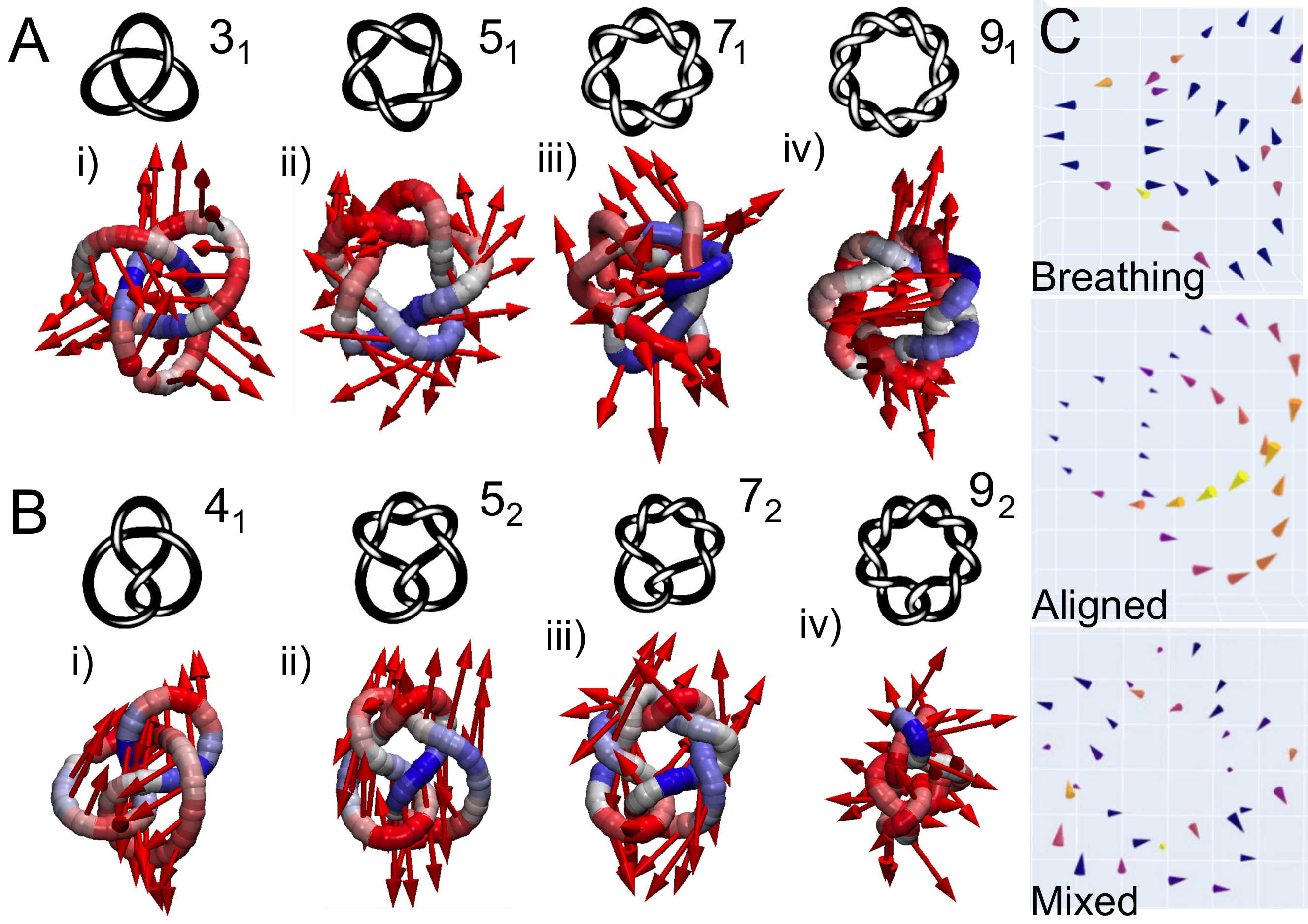}
    \caption{A subset of studied prime knots with respective dominant motion patterns based on principal component analysis of simulated trajectories depicted as arrows. (a) Torus knots $3_1$, $5_1$, $7_1$ and $9_1$. The trefoil knot $3_1$ is also a twist knot. (b) Twist knots $4_1$, $5_2$, $7_2$ and $9_2$. Arrows represent the eigenvectors of the first principal component (PC1) derived from simulation trajectories, and the chain colors represent the amplitude of motion (blue-small to red-large).(c) An example representation of orthogonal, aligned, and mixed motions from simulation trajectories.} 
    \label{fig1}
\end{figure}

Knots are formally defined as closed curves embedded in $\mathbb{R}^3$ that cannot be untangled without a cut. Prime knots are a subset of closed curves that cannot be decomposed, i.e. expressed as a connected sum of two non-trivial knots. Prime knots contain numerous subcategories, and two of those are the focus of this manuscript---torus and twist knots (Fig. \ref{fig1}). These knots have intrinsic symmetries, forming braid-like crossings along a circular path. The only difference between torus and twist knots is that twist knots also contain self-links---loop-like structures that are not present in torus knots (compare Fig. \ref{fig1}(a) and (b)). The notation used throughout the manuscript, such as $3_1$ or $10_{124}$, reflects the crossing number, in this case 3 and 10, respectively. This is the number of times a knot crosses itself, and a higher number of crossings indicates higher complexity. Subscript values 1 and 124 are based on tabulated knot configurations, and are used to identify a knot with the same number of crossings, but different configurations. 

Knots have played an important role in the understanding and controlling the dynamics and mixing of fluids such as liquid crystals~\cite{Smalyukh2020, Alexander2019}, plasma~\cite{Smiet2017, Smiet2019} and even quantum fluids~\cite{Hall2016, Kleckner2016, Barenghi2007}. Topological stability of knots is one of the reasons for the growing interest in the field~\cite{Oberti2016, Schaufelberger2020, Fielden2017}, with applications in information storage from quantum- to biological scales~\cite{Bonesteel2005, Lim2015}. Knots provide an excellent opportunity to study the intricate relationship between topology and dynamics in various physical and biological systems~\cite{Ghrist1998, Boi2005, Michieletto2022, Smrek2021}. It is well known that chain entanglement strongly affects dynamical, thermal and mechanical properties of polymers, including biopolymers such as DNA~\cite{Michieletto2014, Michieletto2017, Signorini2021, Rosa2012}. Knots have been used in nanofluidic applications and nanopore sequencing, serving as invaluable tools for genomic analysis~\cite{Klotz2017, Amin2018}. While many physical properties of knots can be elegantly captured in experiments, the details of the relationship between knotting, entanglement and dynamics are not well understood.

In this work, Molecular Dynamics simulations were used to investigate the link between topology and dynamics of torus and twist knots. Detailed analysis of simulation trajectories revealed that knots undergo `coherent' motions, in which different parts of the knot move cooperatively producing complex motion patterns. Two types of cooperative motions were identified---``breathing" and tangential motions. Principal component analysis allowed us to decompose those dynamics into fundamental ones which we termed orthogonal, tangential, and mixed motions. Based on principal component analysis, free energy landscapes (FEL) and correlation matrices were generated. Together with van Hove correlation functions, they demonstrated how knot complexity can lead to what we call ``topological dynamical arrest", whereby knots with a high number of crossings display frozen-in dynamics purely as a result of topology. Lastly, autocorrelation functions (ACF) of the trajectories were computed, which showed that dynamical trends in decreasing knot size---as knots evolve from completely random dynamics to dynamical arrest---are entirely determined by knot complexity and topology. An attempt to disentangle how various parameters govern and give rise to unique knot dynamics, this work presents a link between the motions, dynamical arrest, knot complexity and topology.

\section{Results and Discussion}\label{sec: results}
\subsection*{Breathing, tangential, and basis motions}\label{sec: results_coop motions}
Polymer chains, entangled or not, experience different types of motions as a result of self-interactions as well as interactions with the solvent. Here, knotted polymer structures are studied, and to identify motions unique to knots of a given subgroup and complexity, principal component analysis was performed. Principal components (PC) represent dominant motions (fluctuations) extracted from simulation trajectories. The strongest of those---PC1 and PC2---can be classified into specific types of motion and used to characterize the overall dynamics of a knot. For this classification, the PC eigenvectors were used as fluctuation vectors (FV) to construct correlation matrices, which display correlations between the root-mean-squared (RMS) fluctuations of individual particles of a knot. The appearance of correlation matrices is closely related to the behavior of FVs. Examining those matrices thus allows comparison of the dynamics represented by different PCs. As a result, three general types of motions were identified---orthogonal, aligned, and mixed (Fig. \ref{fig1})---which underlie the aforementioned breathing and tangential categories of cooperative motions.

\begin{figure}[ht]
    \centering
    \includegraphics[width=\linewidth]{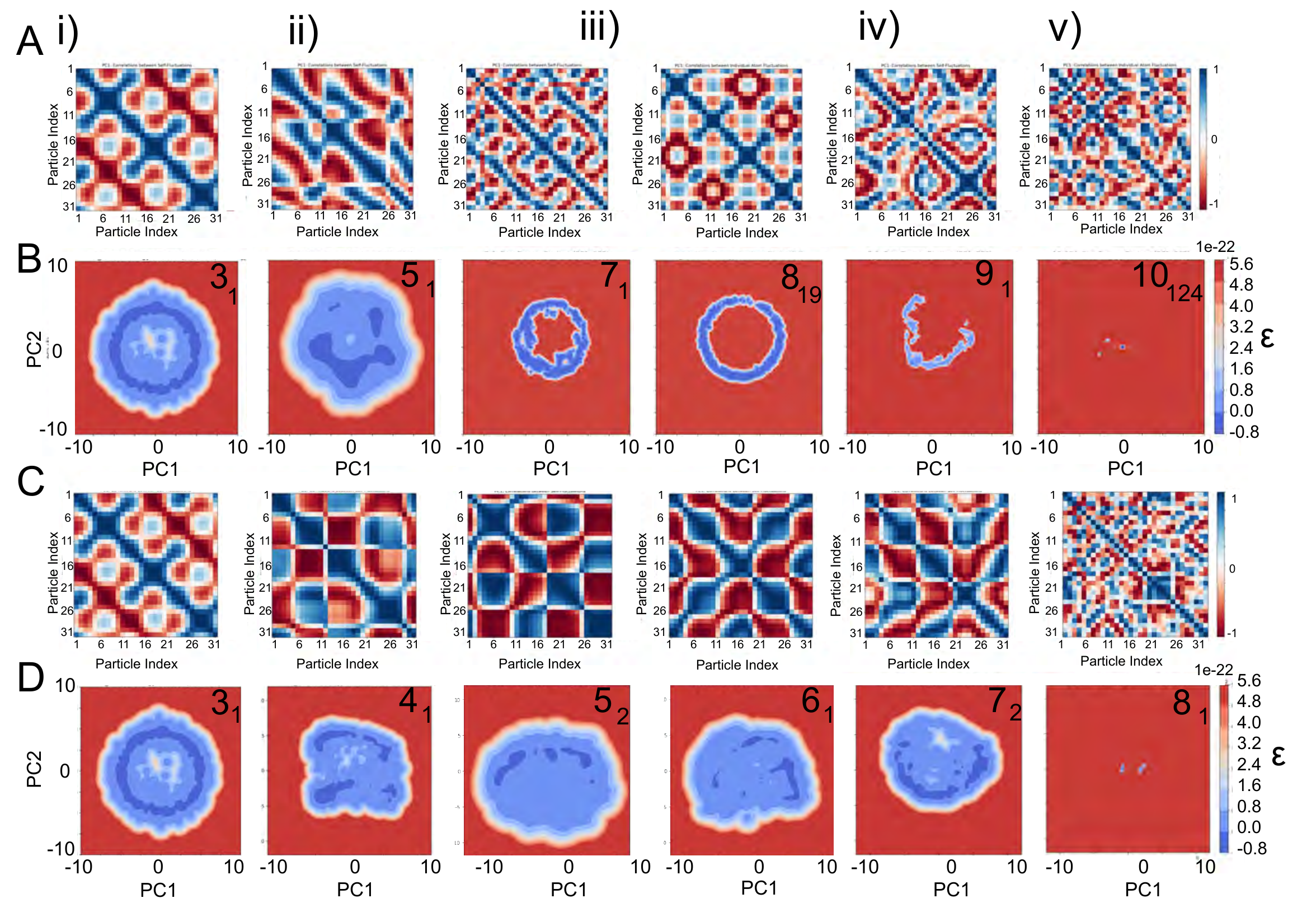}
    \caption{Analysis of torus and twist knot correlation matrices and their respective free energy contour profiles. (a) correlation matrices of torus knots i) $3_1$ ii) $5_1$ iii) $7_1$, iv) $8_{19}$, v)$9_1$, vi) $10_{124}$, showing correlated motion patterns in red and anti-correlated motions in blue. (b) Free energy profiles of torus knots derived from principal component analysis of simulation trajectories. i) $3_1$ ii) $5_1$ iii) $7_1$, iv) $8_{19}$, v)$9_1$, vi) $10_{124}$. Blue regions represent favourable regions of the phase space that are readily explored within a simulation trajectory and red regions remain unexplored. (c) correlation matrices of twist knots i) $3_1$ ii) $4_1$ iii) $5_2$, iv) $6_1$, v)$7_2$, vi) $8_1$, showing correlated motion patterns in red and anti-correlated motions in blue. (d) Free energy profiles of twist knots derived from principal component analysis of simulation trajectories. i) $3_1$ ii) $4_1$ iii) $5_2$, iv) $6_1$, v)$7_2$, vi) $8_1$. Blue regions represent favourable regions of the phase space that are readily explored within a simulation trajectory and red regions remain unexplored.}
    \label{fig2}
\end{figure}

Orthogonal motions are dynamics in which the FVs of individual particles are largely perpendicular to the knot strand. These motions are generally characterized by square patterns in correlation maps as seen in that of $4_1$ (Fig. \ref{fig2}(c,ii)). Such maps show even-sized boxes with alternating blue and red colors, which correspond to positive and negative correlations, respectively. The square patterns may also appear to be smoother as in the correlation map of $3_1$ (Fig. \ref{fig2}(a,i)). Since orthogonal motions generate in- and outward fluctuations of the knot strand, we argue that orthogonal dynamics account for breathing motion, which is the recursive tightening and loosening of the crossing regions in a knot.

Aligned motions, on the other hand, describe motions in which the FVs of adjacent particles point in the same direction, that is, are ``aligned" with each other. These motions in general coexist with smooth patterns in correlation maps where the same color continues in the oblique direction (see $5_1$ in Fig. \ref{fig2}(a,ii)). In low complexity knots, if the geometry permits, aligned and orthogonal dynamics cooperate in propelling tangential motion along the knot, causing the knot strands to smoothly slither or pass by each other. It is aligned motions that give directionality to tangential dynamics while orthogonal motions open up tighter parts of the knot to allow the slithering of the knot strands. Lastly, mixed motions, which are neither orthogonal nor aligned (or a combination of both), are shown in the PC3 FV plot of $5_1$ (Fig. \ref{fig1}(c)).

Often, these three types of general motions are present in knots with varying intensities. The torus knot $3_1$, for instance, is dominated by orthogonal (breathing) motions in its PC1 dynamics, and yet in PC3, the varying orientations of the aligned FVs match the geometry of the knot, describing the knot's tangential dynamics (Fig. \ref{fig1}(c)). This contrasts with $5_1$, in which the hierarchy of motions is reversed, with PC1 describing tangential motions (again, with a proper geometry) and PC3 the breathing. In both $3_1$ and $5_1$, mixed motions appear in fifth or higher PCs. Different compositions of these basis motions create complex patterns of dynamics, giving each knot a unique dynamical signature.

\subsection*{Comparison of torus and twist knots}\label{sec: results_torus-twist comparison}
Torus and twist knots share common characteristics in their dynamics. First, the tighter parts of the knot tend to have tangential motions, while the looser parts tend to exhibit breathing motions. As evident in FV plots (see Fig. S2), they become more tangential to the strand in the vicinity of crossings, and more perpendicular near the loops or self-links of the knots.

As knot complexity increases, knot's motions become more confined due to their complex topology, eventually culminating in dynamical arrest in both types of knots. The higher the crossing number, the narrower the explored basins of FELs, and highly complex knots such as $8_1$ and $10_{124}$ undergo a transition into a state where the structure is too tightly intertwined to experience any motion at all. 

In general, torus knots (Fig. \ref{fig2}(b)) have highly symmetric, circular FELs, while twist knots (Fig. \ref{fig2}(d))---except $3_1$, which is both torus and twist---have asymmetric, broad and level ones. In addition, unlike twist knots, whose correlation maps along the first two PCs show relatively clear square patterns, those of torus knots lack such patterns (compare Fig. \ref{fig2}(a) and (c)).

Torus and twist knots also exhibit different trends as knot complexity increases: For twist knots, the explored areas of the FELs remain relatively constant, as well as the sizes of ``squares" in the correlation maps (see Fig. \ref{fig2}(c) and (d)). This is in contrast to what is observed in torus knots, in which the areas of FELs and the patterns of correlation maps vary significantly. It is apparent that the presence of self-links in twist knots is the source of these different behaviors, demonstrating how even a relatively minor change in knot architecture can lead to considerable changes in their dynamics. 

\begin{figure}[ht]
    \centering
    \includegraphics[width=1.0\linewidth]{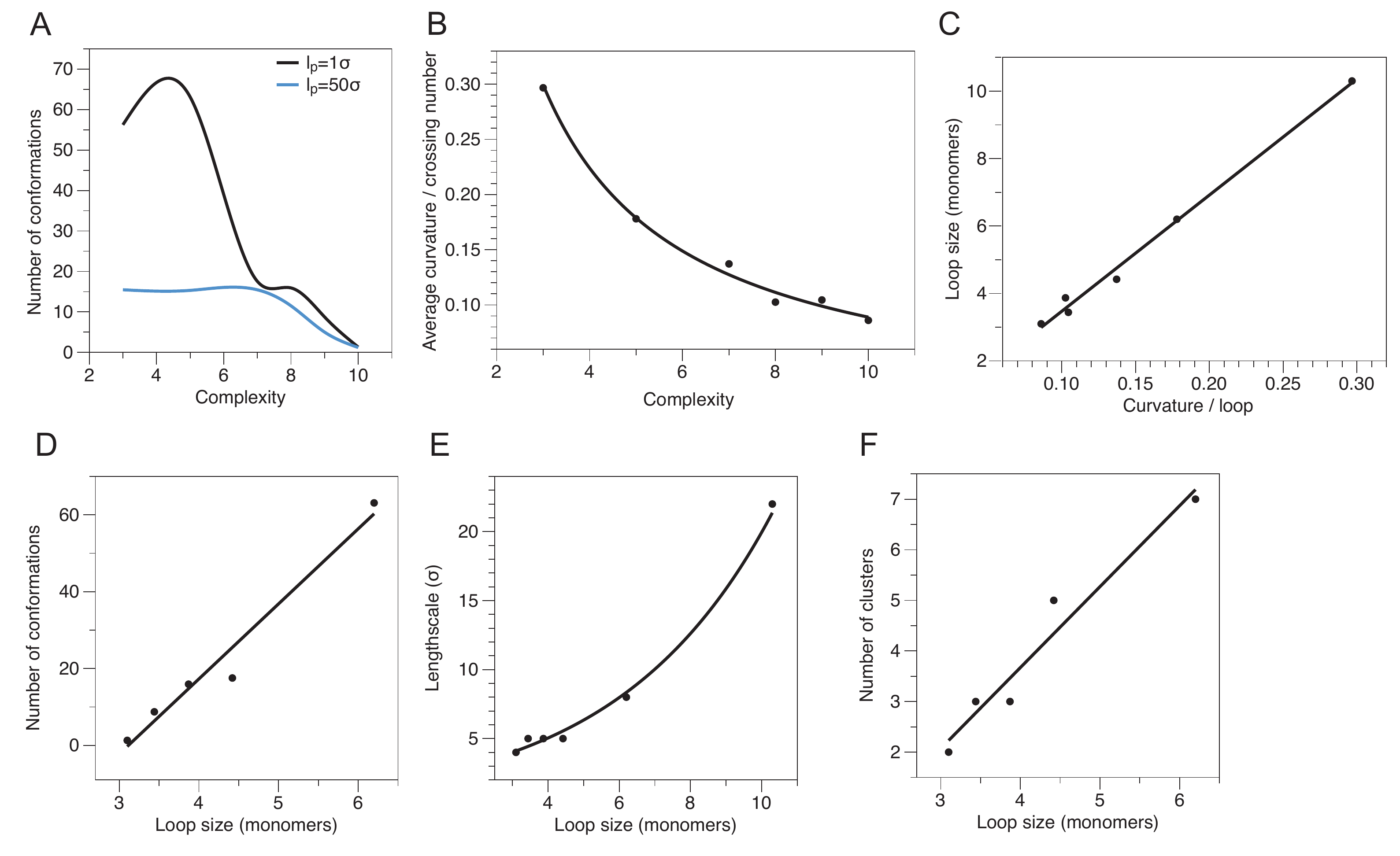}
    \caption{Conformational analysis of knots of varying complexity provides insights into details of their structure and dynamics. (a) The number of accessible conformational states derived from free energy profiles for persistence lengths $(l_p)$ 1 and 50$\sigma$ for knots of varying complexity shows a complex nonlinear relationship. (b) Average curvature / crossing number as a function of complexity follows a power law relationship. (c) Loop size (Number of monomers / number of crossings) as a function of curvature per loop shows a linear relationship. (d) Number of conformations as a function loop size follows a linear relationship. (e) Length-scale grows exponentially with loop size. (f) The number of clusters derived from simulation trajectories as a function of loop size follows a linear curve.}
    \label{fig3}
\end{figure}

\subsection*{The link between knot complexity, motion lengthscales, and dynamical arrest}\label{sec: results_making connection}

Here, the focus is on the connection between structure in general as well as specific structural elements such as loops, and dynamics, and how both respond to topological changes, such as knot complexity. It is evident that knots with a high number of crossings (higher complexity) tend to also be denser-- but are local densities the only factor that contributes to the inevitable dynamical arrest in highly complex knots? The simple answer is it is not-- and while high density does strongly contribute to dynamical arrest, the extent of it also depends on the size of loops, chain flexibility and local geometry. In general, the number of accessible conformations reduces non-linearly as knot complexity increases. (Fig. \ref{fig3}(a), and as expected, the number of accessible states is decreased with an increase in the persistence length of the chain. A careful look into trajectory-averaged curvature per loop as a function of complexity follows a power law, suggesting that highly curved loop segments are mostly present in low complexity knots. In fact, loop size is linearly dependent on curvature per loop, confirming that on average,larger knots have highly curved loops. (Figs. \ref{fig3}(b, c). Perhaps unsurprisingly, this leads to a linear relationship between the number of energetically accessible conformations and loop size. (Fig. \ref{fig3}(d) Fourier transforms of correlation maps were examined in order to study how the correlated motions evolve as a function of knot complexity. For each PC1 correlation map, the largest wavelength (length-scale) of blue-to-red inversion was calculated from the map's Fourier transform
(Fig. \ref{fig3}(e)). As expected in both types of knots, a general decrease in lengthscale was observed as the crossing number increased. This trend stems from the fixed size of knots (all knots are $N=31$ monomers), which results in decreased spacing between crossing regions, or ``length of loops," making the correlation patterns more granular towards higher knot complexity. Twist knots have constant lengthscales from $4_1$ to $7_2$, in agreement with the previous observation that the correlation maps of those four knots have nearly identical sizes of squares. A notable feature is the sudden drop in lengthscale after $7_2$, which contrasts the relatively smooth decrease in lengthscale in torus knots. This reflects the sharp transition into dynamical arrest in twist knots that is observed in their FELs (Fig. \ref{fig2}(d)). Evidently, all these unique characteristics of twist knots---the sharp drop and the flat region in lengthscale as well as the nearly identical correlation maps of the four twist knots---stem from the presence of self-links, which may sustain the variety or strength of dynamics against the decreasing length of loops or give rise to similar dynamics in the four twist knots of different complexities. Interestingly, lengthscale scales exponentially with loop size (Fig. \ref{fig3}(e)).

Cluster analysis of simulation trajectories also illustrate these interesting trends: the number of dynamical clusters decreases as a function of knot complexity, with torus knots following a linear decline and twist knots mostly only having two identifiable clusters. In this context, clusters signify types of observed motions derived by grouping similar basins in the FELs (see Section \ref{sec: methods} for details on cluster analysis). Like the number of conformational states, number of clusters scales linearly with loop size.

\begin{figure}[ht]
    \centering
    \includegraphics[width=1\linewidth]{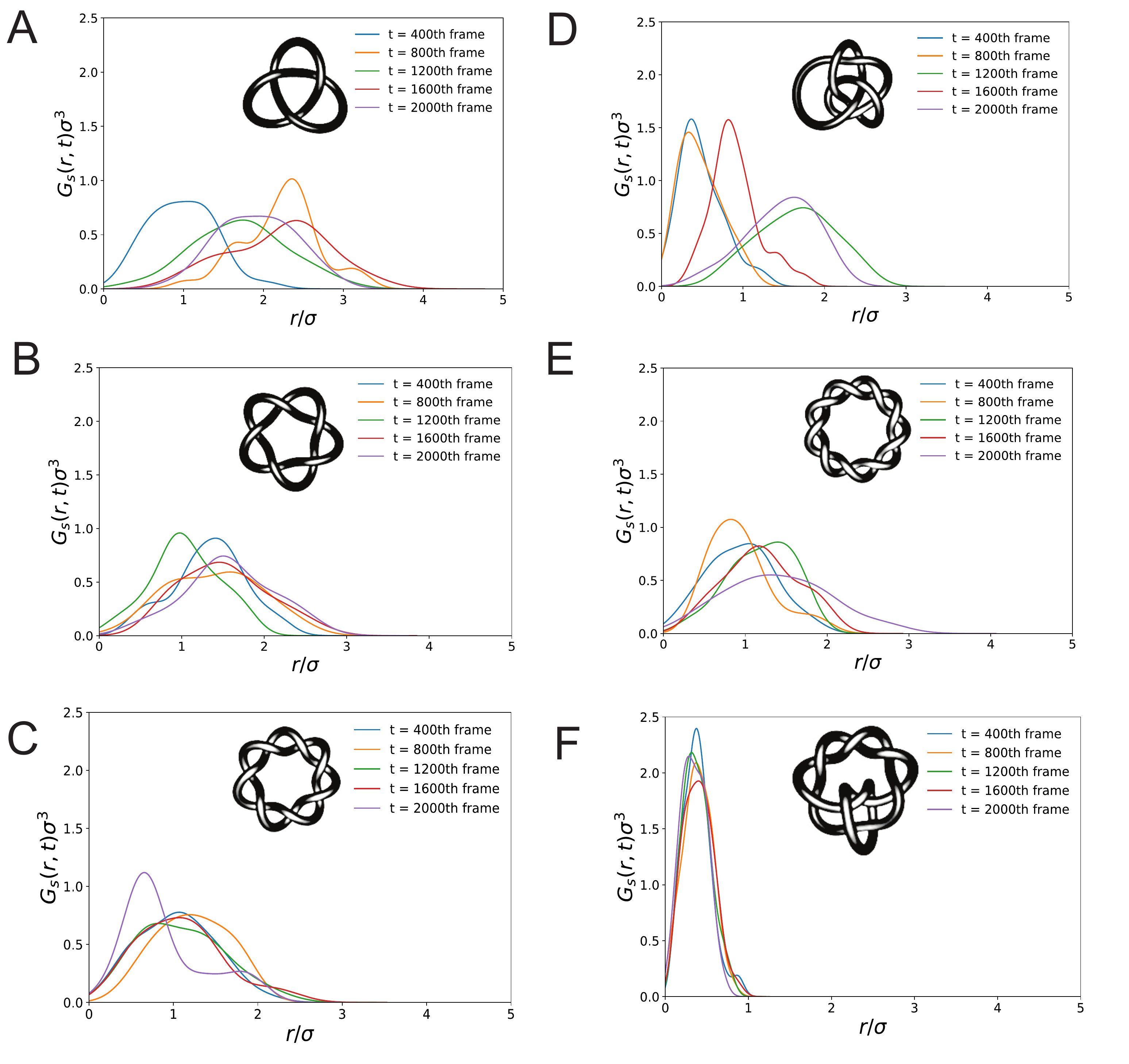}
    \caption{Van Hove self-correlation function for torus knots at different time frames of the simulation trajectory. (a) $3_1$ (b) $5_1$ (c) $7_1$, (d) $8_{19}$, (e)$9_1$, (f) $10_{124}$.}
    \label{fig4}
\end{figure}

\begin{figure}[ht]
    \centering
    \includegraphics[width=1\linewidth]{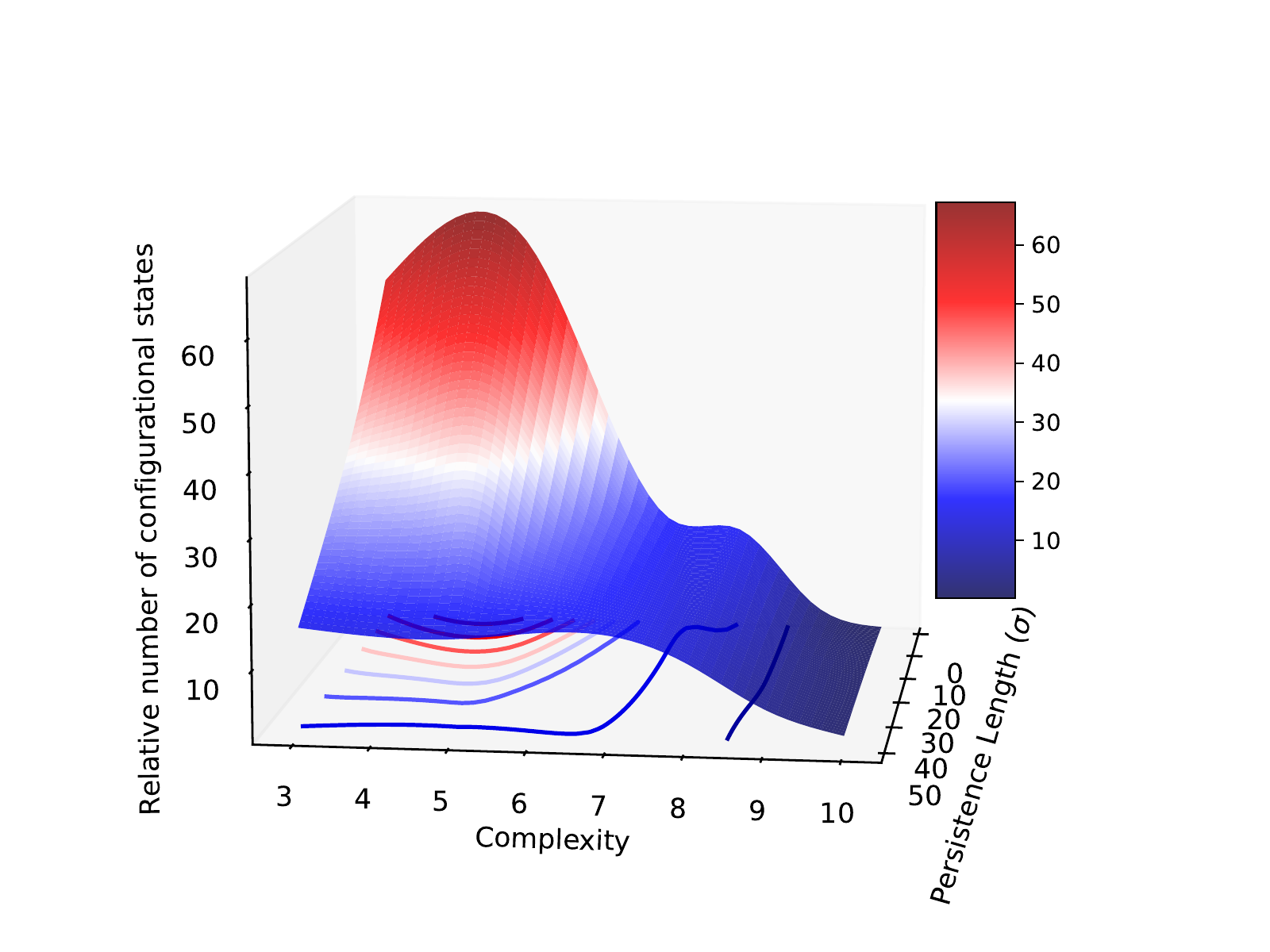}
    \caption{A phase diagram representing a relationship between the number of energetically feasible conformational states, crossing number (complexity) and persistence length. Three major phases are observed: a liquid-like ergodic phase (red), a glass-like dynamically arrested state (blue) and an intermediate state that involves cooperative motions.}
    \label{fig5}
\end{figure}

All the data presented so far reveal the key characteristics of dynamical arrest in high-complexity knots: highly restrictive free energy profiles and small correlation lengthscales. In order to verify if those are indeed marks of dynamical arrest, van Hove self-correlation functions were computed for all knots, and it is clear that the RMS profiles for dynamically arrested knots such as $8_1$, $9_2$, and $10_{124}$ are stable over long times, as shown in Fig.~\ref{fig4}(f), \ref{fig5} and Fig. S2. These findings suggest that topology alone can lead to dynamical arrest in knotted structures.

\subsection*{Controlling randomness: topology-dependent dynamical trends in varying knot size}\label{sec: results_ACF}

\begin{figure}[h!]
    \centering
    \includegraphics[width=0.9\linewidth]{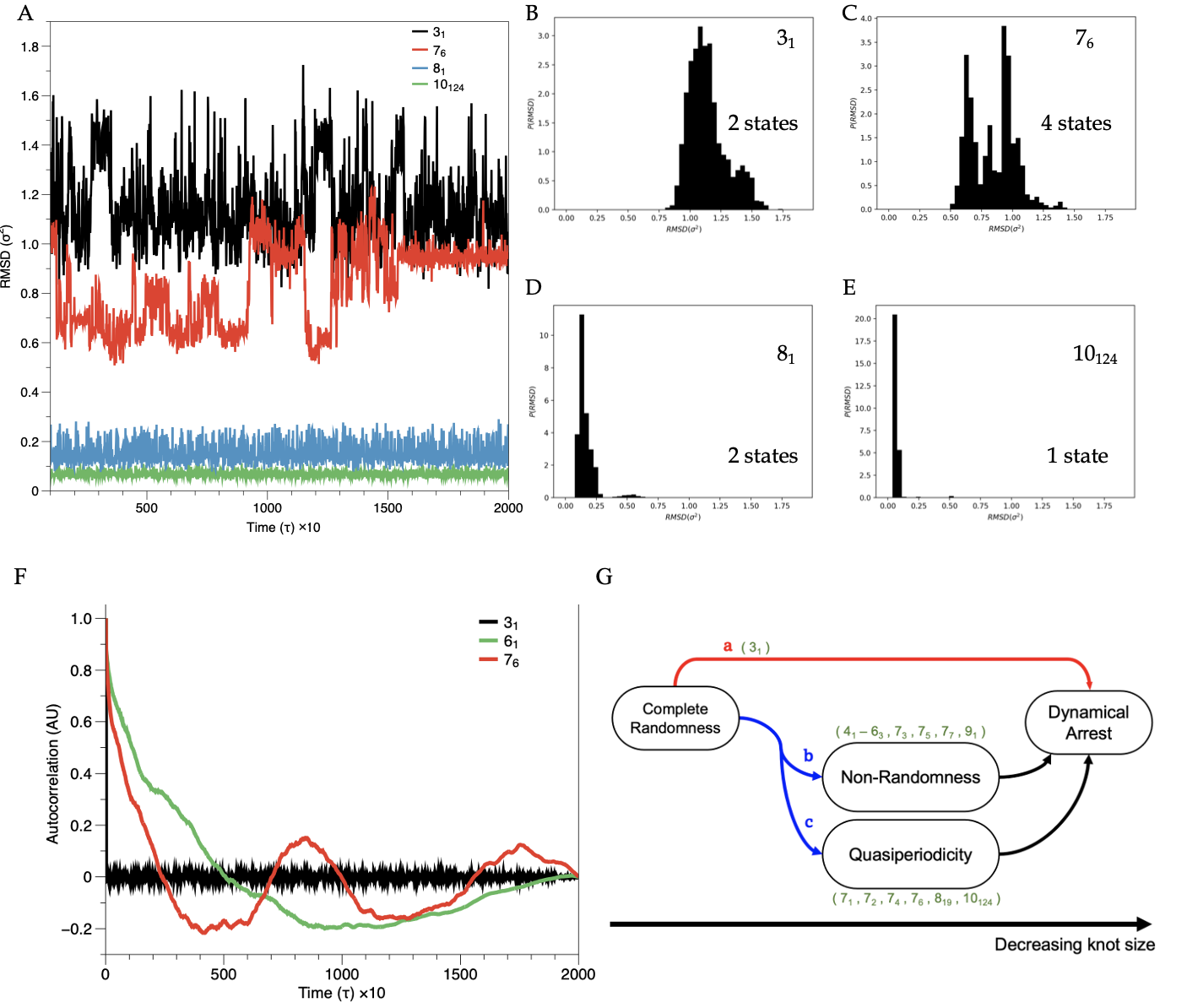}
    \caption{Decoupling density-induced dynamical arrest from geometry-induced one. (a) Spherically confined knots $3_1$ and $7_6$ showed higher RMS displacements as opposed to $8_1$ and $10_{124}$. Knot $7_6$ showed a unique profile that corresponds to different conformational states. (b) A histogram based on RMSD values from (a) for $3_1$, showing at least 2 peaks. (c) As in (b), $7_6$ showing at least 4 peaks. (d) As in (b), $8_1$, showing at least 2 peaks. (e) As in (b), $10_{124}$, showing only one peak.(f) Autocorrelation functions of some knots and the dynamical trends observed in varying knot size. Collective autocorrelation functions of $3_1$ with size $N=11$, of the $6_1$ knot with size $N=27$, and the $7_6$ knot with size $N=31$. Each is an example illustrating complete randomness, non-randomness, and quasiperiodicity, respectively. (g) A schematic of the transition from complete randomness to dynamical arrest as knot size decreases. $3_1$ does not lose randomness until it reaches a state of dynamical arrest (Route Red: a). For knots with crossing number of $>4$, randomness gradually decreases as knot size shrinks and non-random patterns appear. Only a subset of high-complexity knots exhibit quasiperiodicity in their collective ACFs.}
    \label{fig6}
\end{figure}
It may be tempting to assume that dynamical arrest in higher complexity knots is simply triggered by higher contact densities. In order to decouple the effects of local density from geometry, the structures were spherically confined to reach the same effective volume, demonstrating that knots of varying complexities display different dynamics as a result of geometry alone. To confirm this, a number of structures ranging from a linear (unknotted) chain to a knot with 10 crossings were evaluated under spherical confinement to control for density, with the idea  that if density was the only factor influencing the dynamics of knots, root mean squared displacements would be the same for all the different knot topologies. Yet our results (Fig. \ref{fig6}) demonstrate that this is not the case, and knots of higher complexity display a dramatic decrease in mobility, confirming the hypothesis that topology plays a critical role in the initiation of this new kind of purely topology-driven dynamical arrest.  
Autocorrelation analysis of knot trajectories showed interesting dynamical trends as knot size was varied. Autocorrelation functions (ACF) capture deviations from standard random dynamics, and for each knot, both the autocorrelation of the collective structure (``collective ACF," which takes into account all the $3N$ coordinates in the knot) and the autocorrelation of individual coordinates (``individual ACFs") were computed (see Section. \ref{sec: methods} for more detail). The results revealed that randomness in knot dynamics is highly controllable by varying knot size. When the size of a knot is large enough that the entire structure is loose, both the knot's individual and collective ACFs are flat with values nearly equal to zero, implying complete randomness in the knot's dynamics (see Fig. S4---ACFs of $3_1$ $N=31$ and some large high-complexity knot). In the lowest-complexity knot ($3_1$), randomness is sustained regardless of knot size (see $3_1$ in Fig. \ref{fig6}). However, knots with crossing number of 4 or higher begin to diverge from complete randomness as their size decreases and show significant non-random patterns in their trajectories. In some of these higher-complexity knots, periodic trends emerge at some particular knot sizes, which we termed ``quasiperiodicity" Examples of this general non-randomness and quasiperiodicity are show in Fig. \ref{fig6} (a) ($6_1$ and $7_6$) (see S5 for more examples). Quasiperiodicity (Fig. \ref{fig6} (g, Route c) is a subset of general non-randomness (Route B) and characterized by a vanishing semi-sinusoidal curve in the knots' collective ACFs (see Fig. \ref{fig6} (f) $7_6$). Interestingly, it only emerges in some torus knots $7_1$, $8_{19}$, and $10_{124}$ and in certain high-complexity knots such as $7_2$, $7_4$, and $7_6$, which are neither torus nor twist. Upon closer examination, it was noticed that those knots generally have ring-shaped FELs (see Fig. \ref{fig2}(b) and Fig. S6 (FELs of ``other" knots)). A striking feature of these results is that the ``fate" of dynamical trends in varying knot size is solely determined by knot complexity and topology: knots must have an average crossing number of at least 4 or 5 in order to exhibit non-randomness. The type of non-randomness depends on topological structures, with only a subset of high-complexity knots allowed to have quasiperiodic behaviors. This is another demonstration that topology alone can affect the dynamics of entangled objects.
\section{Conclusion}\label{sec: conclusions}
In this work, the dynamics of prime knots was studied, focusing on torus- and twist knots due to their intrinsically symmetrical structures. Results of this work revealed a strong connection between knot dynamics, topology and knot complexity. Dominant motions in torus and twist knots were studied in detail, and three types of motion patterns were identified: ``breathing", aligned, and a combination of the two. These motions work in unison to generate complex patterns in the dynamics of individual knots. It was also found that randomness in knot dynamics can be modulated by varying knot size. As knot size shrinks, knots exhibit either complete randomness, non-randomness, or quasiperiodicity depending on knot complexity and topology, before reaching a state of dynamical arrest. It was observed that some knots undergo cooperative motions as they reach higher complexity, uniquely modulating conformational patterns of a given knot. Intriguingly, knots of high complexity were found to undergo a so called topology-driven dynamical arrest that manifests in a dramatic reduction of explored configuration space accompanied by a shift in typical patterns of motion in knots. Understanding periodic and non-random dynamics as well as topology-guided dynamical arrest will enable future studies of entangled systems not limited to knots, leading to the control and regulation of local dynamics of entangled structures, including biological systems such as proteins, RNA and chromosomes.  
\section{Methods}\label{sec: methods}
Initial configurations of prime knots (Fig. \ref{fig1}(a)) were generated based on models from KnotPlot software \cite{SchareinPhD}. Coordinates were interpolated using three-dimensional parametric curves in order to build knots of desired size. Each curve was scaled to fit 31 particles (a number that produces stable simulations for all studied knots), and along the curves knots were constructed particle by particle, ensuring that all bond lengths are within $\pm0.08\sigma$ and the inter-atomic distances between non-bonded pairs at least $0.72\sigma$, where $\sigma$ is the monomer radius in Lennard-Jones units.
Molecular Dynamics simulations described in this work were performed using LAMMPS software \cite{LAMMPS}. From the simulation trajectories, for each knot the data matrix $X$ of shape [$3N, m$] was obtained, where  $N$=31, and $m$ is the number of frames in the trajectory. The transitions and rotations along the trajectory were removed before $X$ was constructed, as the focus was on the internal dynamics of the knots \cite{David2014}. 
Principal component analysis of the data matrix was performed to extract the most dominant fluctuations from Molecular Dynamics (MD) trajectories. The correlation matrix was first calculated
\begin{eqnarray}
    Q = \frac{1}{m}\overline{X}\overline{X}^T
    \label{eq1},
\end{eqnarray}
where $\overline{X}$ is the mean-centered data matrix obtained by subtracting means from the rows of $X$. Eigendecomposition of $Q$
\begin{eqnarray}
    Q = V\Lambda V^{-1}
    \label{eq2}
\end{eqnarray}
yields the orthonormal eigenvectors in the columns of $V$ and the eigenvalues in the diagonal of $\Lambda$.
The principal components (PC) are these eigenvectors $\bm{v_\mathrm{i}}$ and each describe the direction of a particular deviation from the average conformation. Each of their corresponding eigenvalues $\lambda_i$ indicates the portion of the total variance of $X$ that is explained by that PC. A large eigenvalue means that the direction of that specific deviation is frequently explored over the entire trajectory.
Thus, by sorting the PCs in a descending order of their eigenvalues and analyzing the first few dominant PCs, one can identify the strongest motions that are present in the knots. In this work the PCs are referred to as PC1 (the first PC), PC2 (the second), and so forth. Projections along principal components, free energy landscapes, correlation maps, and fluctuation vector plots were examined to characterize the motions of the knots. Cluster analysis, van Hove self-correlation functions, and autocorrelation functions were used in studying the relationship between those motions, knot complexity, and dynamical arrest.
Fig. S2 shows the projections of conformations on the plane formed by two PCs. Each projection contains $m$ scores (shown as dots), and each score provides information of how much the two PCs account for that particular conformation. Values of $z_{ij}$ are obtained by taking the dot product of the $i$th frame and $j$th eigenvector:
\begin{eqnarray}
    z_{\mathrm{ij}} = \bm{x_\mathrm{i}}\cdot\bm{v_\mathrm{j}}
    \label{eq3},
\end{eqnarray}
where $\bm{x_\mathrm{i}}$ is the $i$th column of $\overline{X}$, and $\bm{v_\mathrm{j}}$ the $j$th column of $V$. The calculation of scores can be easily achieved via matrix algebra:
\begin{eqnarray}
    Z = \overline{X}^T V
    \label{eq4}.
\end{eqnarray}
In this case, the scores are the elements of the score matrix $Z$.
Free energy landscapes (FEL) were computed from the distributions of projections along principal components based on the formula
\begin{eqnarray}
\Delta G(r_1, r_2) = -k_\mathrm{B} T \ln{P(r_1, r_2)}
\label{eq5},
\end{eqnarray}
where $k_\mathrm{B}$ is the Boltzmann constant, $T$ the temperature of the simulation, and $P(r_1, r_2)$ the estimated probability density function along the axes $r_1$ and $r_2$ of some two PCs. $P(r_1, r_2)$ was estimated (Fig. \ref{fig1}(c)) by fitting a score plot using a two-dimensional Gaussian kernel function with the improved Sheather Jones bandwidth~\cite{KDEpy}. 
After obtaining $P(r_1, r_2)$, one can plot the three-dimensional free energy landscapes and also their free energy contour maps.
Correlation maps display the correlations between the root-mean-squared (RMS) fluctuations of individual particles in a knot. The correlation between particles $i$ and $j$ is defined as
\begin{eqnarray}
    c_{\mathrm{ij}} = \frac{<\Delta\bm{R_\mathrm{i}}\cdot\Delta\bm{R_\mathrm{j}}>}{\sqrt{<\Delta\bm{R_\mathrm{i}}^2>\cdot<\Delta\bm{R_\mathrm{j}}^2>}}
    \label{eq6},
\end{eqnarray}
where $\Delta\bm{R_\mathrm{i}}$ and $\Delta\bm{R_\mathrm{j}}$ are the fluctuation vectors. These correlations $c_{ij}$  comprise the $N\times N$ correlation map $C$. Since the directions of dominant conformational changes are described by the PCs, one can use the PC eigenvectors as the fluctuation vectors $\Delta\bm{R}$ of the particles. Thus, the correlation map $C$ of each PC can be found, and for this calculation the algorithm developed by Eyal et al.\cite{ANM} was used. Visualizing these correlations as heat maps provides a means to examine and compare the specific correlated motions described by different PCs.
Fluctuation vector (FV) plots show the PC eigenvectors superimposed onto initial conformations. By examining how the fluctuation vectors are aligned with the knot strands, one can determine the types of dynamics the PCs represent, such as orthogonal, aligned, and mixed motions. One can also deconstruct each PC to study its contribution to the overall cooperative dynamics---tangential and breathing. This can be achieved by projecting the PC eigenvectors onto the Frenet frame $(\mathbf{T}, \mathbf{N}, \mathbf{B})$ along the knot, where the tangential vector $\mathbf{T}$ describes tangential dynamics, and the normal $\mathbf{N}$ and binormal $\mathbf{B}$ vectors breathing dynamics. To reproduce the overall motions along a particular Frenet vector, one can use equations of the form
\begin{eqnarray}
    \bm{v}_\mathrm{fluc} = \sum_{i} \bm{v_\mathrm{i}}\cdot\bm{\mathrm{F}} \sqrt{\lambda_\mathrm{i}}
    \label{eq7}.
\end{eqnarray}
Here, $\sqrt{\lambda_\mathrm{i}}$ ---the square root of the $i$th eigenvalue---describes the standard deviation along the corresponding PC axis, whose direction is given by the eigenvector $\bm{v_\mathrm{i}}$. The Frenet vector $\mathbf{F}$ should be replaced by either $\mathbf{T}$, $\mathbf{N}$, or $\mathbf{B}$.
This Frenet method is especially useful if cooperative motions depend on more than one basis motion types (e.g. tangential dynamics are due to both orthogonal and aligned motions). The correlation maps for breathing dynamics obtained from this method do generally exhibit square patterns, in agreement with the results obtained from individual PCs (Fig. \ref{fig2}(c)). Such comparison, however, cannot be made between the correlation maps of aligned and tangential dynamics, as the former dynamics do not always match the latter.
Cluster analysis was used to group motions of a knot into clusters based on their similarities. We clustered frames in the simulation trajectory, i.e. the columns of the mean-centered data matrix $\overline{X}$, by using the average-linkage algorithm. We specifically chose this algorithm among other options such as complete linkage or K-means, since average linkage produces more consistent results for MD trajectories. For the clustering criterion, correlations between trajectory frames were used, and for determining the number of clusters, the so-called ``elbow method" was employed. Clusters can be plotted on the PC1-PC2 plane to visualize their locations in phase space, which is achieved by coloring scores in the score plot based on cluster indices. The number of identified clusters serves as a measure of the diversity of motions in a knot, allowing the study of how a certain parameter such as knot size and bond stiffness affects the knot's dynamics. In this work, this number was used to demonstrate the link between knot complexity and dynamical arrest: knots become tighter as the crossing number increases, which confines knot dynamics to fewer conformations. This is exactly reflected in their lower numbers of identified clusters, signifying less variety in their motions.
The van Hove correlation function gives the average probability density of finding a particle at position $\mathbf{r'}+\mathbf{r}$ at time $t'+t$, given that there was a particle at position $\mathbf{r'}$ at time $t'$. In other words, it describes the correlation between the ``presence" of a particle in some position and the presence of another (or the same) particle in some position at some reference time~\cite{vanHove}. Normally the reference time $t'$ is taken to be zero. For a system of $N$ particles, the van Hove correlation function can be written as
\begin{eqnarray}
    G(\mathbf{r},t) = \frac{1}{N} \langle \sum_{i=1}^{N}\sum_{j=1}^{N} \delta^{(3)}(\mathbf{r}-\mathbf{r_i}(t)+\mathbf{r_j}(0)) \rangle
    \label{eq8},
\end{eqnarray}
where $\mathbf{r_i}$ and $\mathbf{r_j}$ represent particle positions, $\delta^{(3)}(\cdot)$ a three-dimensional Dirac delta function, and $\langle\cdot\rangle$ an ensemble average (an average over identical systems). At time $t=0$, the function reduces to
\begin{align}
    G(\mathbf{r},0) &= \frac{1}{N} \langle \sum_{i=1}^{N}\sum_{j=1}^{N} \delta^{(3)}(\mathbf{r}-\mathbf{r_i}(0)+\mathbf{r_j}(0)) \rangle\\
    &= \delta^{(3)}(\mathbf{r})+\frac{1}{N} \sum_{i\neq j}\langle\delta^{(3)}(\mathbf{r}-\mathbf{r_i}(0)+\mathbf{r_j}(0)) \rangle\\
    &= \delta^{(3)}(\mathbf{r})+g(\mathbf{r})
    \label{eq9},
\end{align}
where $g(\mathbf{r})$ is the usual radial distribution function~\cite{vanHove}. In the case of systems composed of distinguishable particles, the function $G$ can be separated into two parts:
\begin{align}
    G(\mathbf{r},t) &= \frac{1}{N} \langle \sum_{i=1}^{N} \delta^{(3)}(\mathbf{r}-\mathbf{r_i}(t)+\mathbf{r_i}(0)) \rangle + \frac{1}{N} \langle \sum_{i\neq j}^{N} \delta^{(3)}(\mathbf{r}-\mathbf{r_i}(t)+\mathbf{r_j}(0)) \rangle\\
    &:=G_s(\mathbf{r},t) + G_d(\mathbf{r},t)
    \label{eq10}.
\end{align}
The ``self" part $G_s(\mathbf{r},t)$ gives the distribution of displacements of each particle from their initial position over time, and the ``distinct" part $G_d(\mathbf{r},t)$ describes the distribution of relative displacements between two different particles at different times. Both $G_s$ and $G_d$ can be calculated from the histograms of corresponding displacements obtained from MD trajectories. In this work, Gaussian kernel density functions with improved Sheather Jones bandwidth were used to fit the histograms and approximate smooth $G_s$ and $G_d$ functions~\cite{KDEpy}. As discussed in Section ~\ref{sec: results_making connection}, the self part $G_s$ (Fig. \ref{fig4}) demonstrated dynamical arrest in high-complexity knots.

The autocorrelation function (ACF) is a function of delay or lag $\tau$ that gives the correlation between a time series $X$ and a delayed copy of itself:
\begin{eqnarray}
    R(\tau) = \frac{\sum_{t=1}^{m-\tau} (X(t)-\overline{X})(X(t+\tau)-\overline{X})}{\sum_{t=1}^{m} (X(t)-\overline{X})^2},
\end{eqnarray}
where $m$ is the total time over which $X$ was recorded (which in our case is the number of frames in a MD trajectory), and $\overline{X}$ is the time-average of $X$. In essence, the ACF is a measure of the self-similarity of a data set over different delay times $\tau$~\cite{ACF_concept}. The magnitude and shape of ACF allow one to infer the dynamical trend of the data, i.e. whether the data is random, periodic, autoregressive, etc. An essentially random time series like white noise has an ACF that is equal to zero for all lags $\tau>0$. For this reason, the standard error of the white noise is used to determine whether a given time series $X$ has statistically significant autocorrelation values~\cite{ACF_concept}. $X$ is said to have a significant autocorrelation value at lag $\tau$ if $R(\tau)$ lies outside the confidence interval given by
\begin{eqnarray}
    \pm \frac{\sqrt{2} z_p}{\sqrt{m}},
\end{eqnarray}
where $z_p$ is the cumulative distribution function of the standard normal distribution (also known as the error function), and $p=0.95$ ($95\%$) the confidence level, i.e. the probability with which a new normal random variable will lie within the interval. Autocorrelation values that fall outside this interval suggest non-randomness in the data. In this work, two kinds of ACFs were computed for each knot, which we named a ``collective" ACF and ``individual" ACFs. A collective ACF is autocorrelation that takes into account all $3N$ coordinates in a knot (each coordinate serves as a time series). The collective ACF is useful in that it encodes the dynamical behavior of the entire knot; however, as it is the ACF of a multivariate data set, it is tricky to determine the confidence interval for the data. Thus, individual ACFs (the ACFs of individual coordinates) were computed in order to verify if an interesting trend shown in the collective ACF is indeed statistically significant. Also, for completeness, ACFs of the other prime knots were computed (not only torus or twist knots), such as $5_2$, $6_3$, $7_2$, etc. 
\bibliography{manuscript}
\end{document}


\title{Supplementary Material: Cooperative Motions and Topology-Driven Dynamical Arrest in Prime Knots} 

\author{Hyo Jung Park}
\affiliation{Smith College}
\author{Lakshminarayanan Mahadevan}
\author{Anna Lappala}
\email{lappala@molbio.mgh.harvard.edu}
\affiliation{Harvard University}

\date{\today}
\maketitle
\begin{abstract}
This document provides supplementary material for the main manuscript ``Cooperative Motions and Topology-Driven Dynamical Arrest in Prime Knots."
\end{abstract}

\section{First Section}\label{sec: intro}

\begin{figure}[ht]
    \centering
    \includegraphics[width=0.7\linewidth]{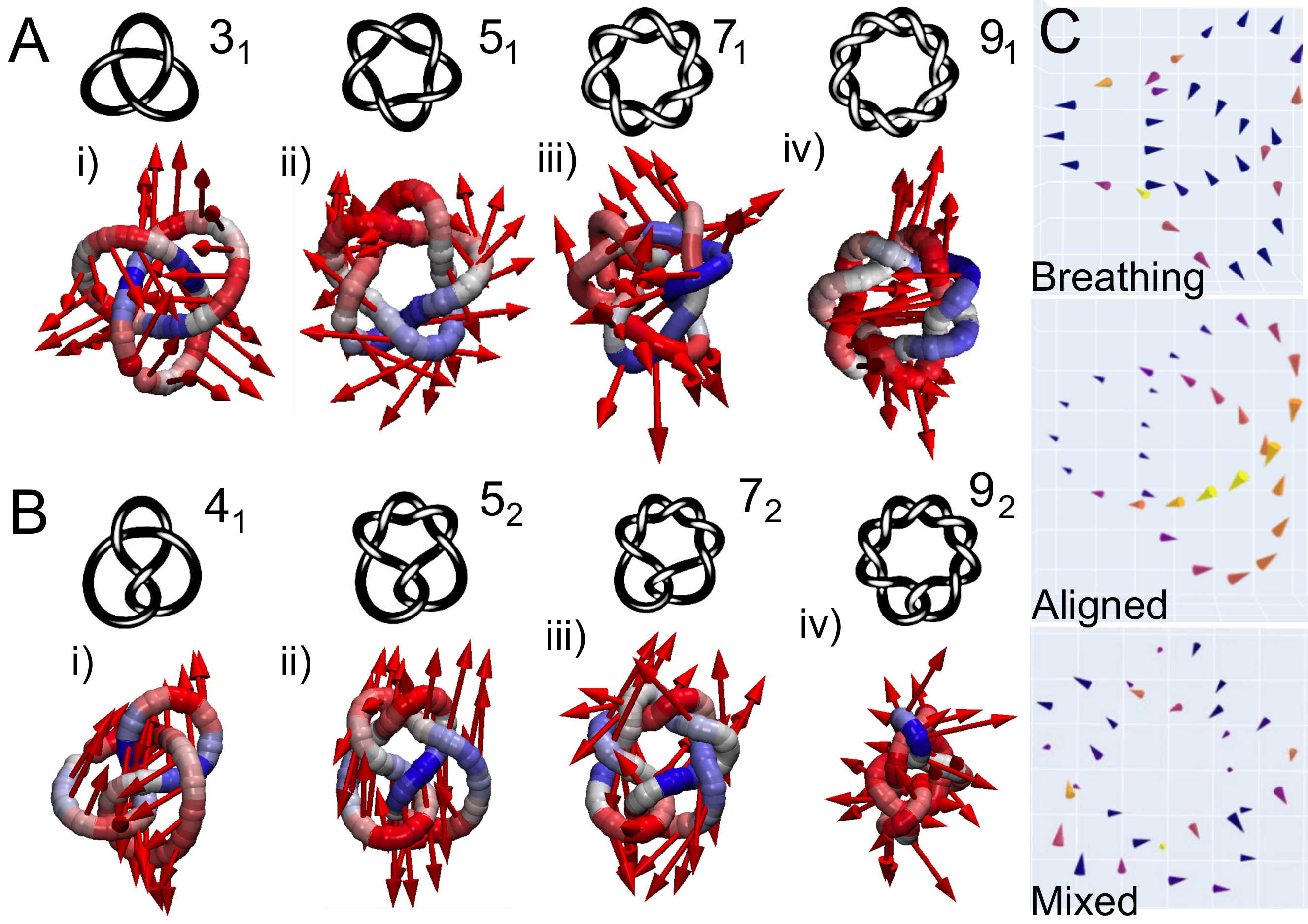}
    \caption{This is first supplementary figure.}
    \label{fig1}
\end{figure}

Refer to figures in this format: (Fig.\ref{fig1}).

